# VAE-QWGAN: Addressing Mode Collapse in Quantum GANs via Autoencoding Priors

Aaron Mark Thomas[1*], Harry Youel[2,3] and Sharu Theresa Jose[1]

[1*]School of Computer Science, University of Birmingham, Birmingham, B15 2TT, United Kingdom.
[2]London Centre for Nanotechnology, University College London, London, WC1H 0AH, United Kingdom.
[3]Department of Physics and Astronomy, University College London, London, WC1E 6BT, United Kingdom.

*Corresponding author(s). E-mail(s): amt326@student.bham.ac.uk;
Contributing authors: harry.youel.19@ucl.ac.uk; s.t.jose@bham.ac.uk;

**Abstract**

Recent proposals for quantum generative adversarial networks (GANs) suffer from the issue of mode collapse, analogous to classical GANs, wherein the distribution learnt by the GAN fails to capture the high mode complexities of the target distribution. Mode collapse can arise due to the use of uninformed prior distributions in the generative learning task. To alleviate the issue of mode collapse for quantum GANs, this work presents a novel **hybrid quantum-classical generative model**, the VAE-QWGAN, which combines the strengths of a classical Variational AutoEncoder (VAE) with a hybrid Quantum Wasserstein GAN (QWGAN). The VAE-QWGAN fuses the VAE decoder and QWGAN generator into a single quantum model, and utilizes the VAE encoder for data-dependant latent vector sampling during training. This in turn, enhances the diversity and quality of generated images. To generate new data from the trained model at inference, we sample from a Gaussian mixture model (GMM) prior that is learnt on the latent vectors generated during training. We conduct extensive experiments for image generation QGANs on MNIST/Fashion-MNIST datasets and compute a range of metrics that measure the diversity and quality of generated samples. We show that VAE-QWGAN demonstrates significant improvement over existing QGAN approaches.

# 1 Introduction

Advances in quantum technology have marked the onset of the Noisy Intermediate Scale Quantum (NISQ) era of quantum computing [1]. This has catalyzed the field of quantum machine learning (QML) [2], which aims to harness the power of quantum computing to enhance learning from classical (e.g., images) and quantum (e.g., arising from quantum sensing) data with the hope of achieving practical advantages over classical machine learning. Within this domain, quantum generative learning (QGL) has emerged as a highly promising avenue of research. Quantum generative learning uses quantum models, implementable on NISQ devices, to learn the unknown data distribution underlying the observed classical/quantum data, with the goal of generating high-quality synthetic samples from the learnt distribution [3].

Quantum Generative Adversarial Networks (QGANs) have emerged as a class of QGL models that are adept at learning complex discrete [4] or continuous data distributions [5, 6]. Analogous to classical GANs [7], QGAN architecture comprises of two components: a *generator* that maps latent random vectors sampled from a prior distribution to synthetic data samples, and a *discriminator* that distinguishes true data samples from synthetic samples. Existing proposals for QGANs include fully quantum models [8, 9], where both the generator and discriminator are quantum, or hybrid models, with a quantum generator and a classical discriminator [10]. Despite promising applications in quantum state generation [11] and quantum state loading [4], QGANs encounter notable challenges, particularly when dealing with high-dimensional classical datasets. Specifically, the number of qubits required to effectively load such classical data into a quantum state often far exceeds the qubit availability in current NISQ devices. Hybrid QGAN models that blend quantum and classical computational paradigms have demonstrated particular promise in this regard.

For the problem of (high-dimensional) image generation, Huang et. al [12] introduce a hybrid *patch-QGAN* that uses a quantum generator and a classical discriminator. The generator consists of multiple sub-generators that each produce a small patch of the output image. The concatenation of all these patches result in the final synthetic sample. The patch-QGAN has been experimentally demonstrated on superconducting quantum processors for the generation of hand-written digits. The key novelty of this framework is the use of multiple quantum sub-generators to synthesize patches of the final output sample. This eliminates the need to directly encode high-dimensional data into a single large-qubit register, a significant challenge in the NISQ era.

The patch-QGAN framework has since been extended to patch-quantum Wasserstein GAN (PQWGAN) [10], which optimizes the 1-Wasserstein distance as the training objective. PQWGAN demonstrates comparable performance to classical GANs while using significantly fewer parameters to generate high-dimensional samples from MNIST and Fashion-MNIST datasets. However, PQWGAN suffers from two key issues: low-quality samples and mode collapse, where the model fails to generate diverse images within the same class. Recent efforts have tried to address these challenges by instead directly operating on a lower-dimensional latent space. For example, reference [13] employs dimensionality reduction via principal component analysis (PCA) first, and then uses a quantum generator to learn these independent and orthogonal vectors of the data. A reverse PCA then reverts the data back to the true dimensional space. Along these lines, reference [14] uses a quantum generator to learn the latent vectors output by the encoder of a pre-trained classical AutoEncoder (AE). With the learning taking place in the latent space, the generated synthetic latent vectors are then mapped to the true data space through the AE's decoder. While these approaches improve mode coverage by focusing on latent spaces rather than pixel spaces, they inherently depend on data pre-processing steps which, in some cases, requires the training of a fully classical generative model.

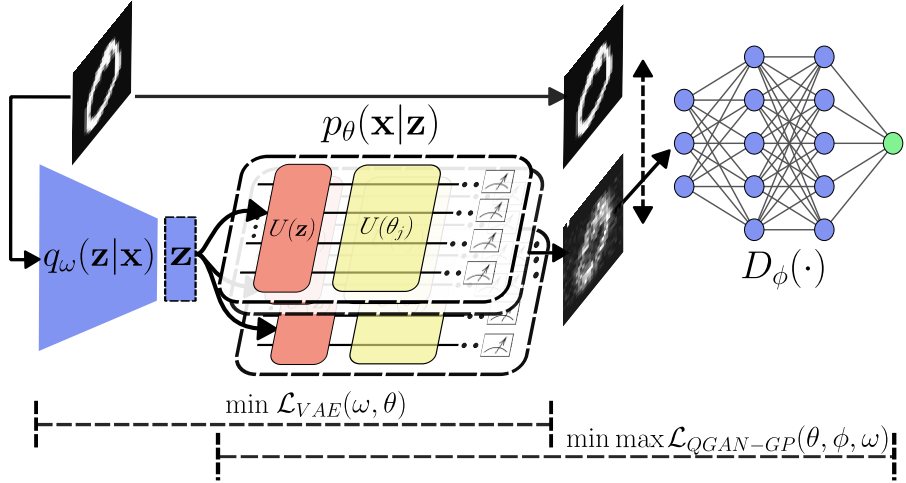

**Fig. 1**: Illustration of the VAE-QGAN Framework. The encoder $q_\omega(\mathbf{z}|\mathbf{x})$ maps input samples $\mathbf{x}$ to a latent code $\mathbf{z}$. The latent vector $\mathbf{z}$ is then encoded into several patches of the quantum generator $G_\theta(\mathbf{z})$ which in turn serves as the VAE's decoder $p_\theta(\mathbf{x}|\mathbf{z})$, generating fake data by reconstructing the input sample. A discriminator $D_\phi$ provides adversarial feedback to improve reconstruction quality and encourage sample diversity. The model is jointly trained using both the VAE reconstruction loss and the WGAN-GP adversarial loss, enabling higher-fidelity generation with reduced mode collapse.

For classical GANs, there has been recent efforts to mitigate mode collapse using variety of approaches ranging from from regularization strategies to constraining the latent space of a model. For instance, DynGAN [15] identifies collapsed samples generated by the model and trains a dynamic conditional generator on a partitioned dataset, progressively recovering missing modes. Several methods focus on enhancing the latent space by learning more informative priors. In VEEGAN [16], an encoder network is introduced to approximately invert the generators mapping from the latent space in an attempt to enforce injectivity and thereby penalising redundant and similar samples. In a similar manner, references [17, 18] both use an encoder in conjunction with a VAE objective to tailor the prior latent distribution to the dataset in question, significantly improving image generation.

Motivated by the above works that explore conditioning of latent space to improve mode collapse, this paper proposes a novel Variational AutoEncoder - Quantum Wasserstein Generative Adversarial Network (VAE-QWGAN) to improve the image generation capabilities of NISQ-based quantum GANs that operate **directly on the pixel space**. Our approach integrates a classical VAE with a hybrid quantum Wasserstein GAN to autoencode a data-informed prior for the QGAN. Unlike [13] and [14], our framework does not require downscaling/pre-processing of input images. We utilize the VAE encoder for sampling structured, data-dependent latent vectors during training, which are then mapped to synthetic images via the quantum generator. This approach can be viewed as extending QGANs with a variational prior that ensures the latent manifold is closely aligned to that of the true data manifold. Consequently, we significantly reduce mode collapse and enhance the fidelity of the generative model. Our key contributions are as follows:

- We propose a novel VAE-QWGAN model, that effectively integrates classical VAE architectures with hybrid quantum GAN methodologies for high-resolution image generation by directly operating on the pixel space. Specifically, VAE-QWGAN fuses the VAE decoder and QGAN generator into a single quantum model with shared parameters and use the VAE's encoder to sample latent random vectors for the QGAN generator during training.
- For inference with VAE-QWGAN, we fit a Gaussian mixture model (GMM) on the latent vectors generated during training. The resulting GMM then serves as the prior distribution during inference, which generates latent vectors that are mapped to synthetic images. The GMM therefore captures the intrinsic characteristics of the data latent space.
- We conduct extensive numerical experiments on MNIST and Fashion-MNIST datasets, comparing our VAE-QWGAN with standard classical and quantum benchmark models. Our experiments demonstrate that VAE-QWGAN better learns the relevant latent space of the input data, leading to improved mode coverage and sample quality relative to the currently established state-of-the-art hybrid quantum modes that operate on the full pixel space. We also compare our results against a classical GAN baseline with significantly more trainable parameters and find that VAE-QWGAN achieves comparable sample quality despite its considerably lower model parameters.

## 2 Background and Preliminaries

In this section, we introduce the building blocks of our model – GANs, VAEs and variational quantum circuits.

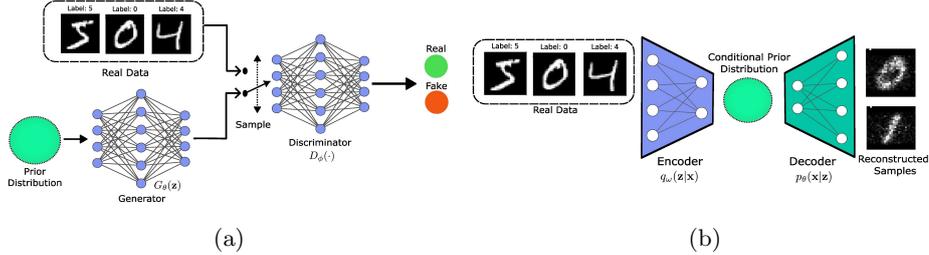

(a)             (b)

**Fig. 2**: Diagrams depicting different generative models referred to in this work **(a)** Classical GAN and **(b)** Classical VAE.

## 2.1 Generative Adversarial Networks

GANs are a class of generative learning models that aim to learn the unknown data distribution $p_r(\mathbf{x})$ that underlies an observed dataset $\mathcal{D} = (\mathbf{x}_1, \ldots, \mathbf{x}_n)$ of $n$ feature vectors [7]. Here, $p_r(\mathbf{x})$ denotes the probability distribution defined over the data space $\mathcal{X}$ with $\mathbf{x} \in \mathcal{X}$. A typical GAN architecture comprises of two components: a *generator* and a *discriminator*. The generator is defined via the function $G_\theta : \mathcal{Z} \to \mathcal{X}$, parametrized by $\theta \in \Theta$, that maps an input latent vector $\mathbf{z} \in \mathcal{Z}$ to the data space $\mathcal{X}$ as $\mathbf{x}' = G_\theta(\mathbf{z})$. The latent space $\mathcal{Z}$ is endowed with a prior probability distribution $p(\mathbf{z})$. The discriminator is similarly defined by a function $D_\phi : \mathcal{X} \to \mathbb{R}$, parameterized by $\phi \in \Phi$, that aims to distinguish between true data sample $\mathbf{x}$ and the generated sample $\mathbf{x}'$ based on the score $D_\phi(\mathbf{x})$ and $D_\phi(\mathbf{x}')$. Together, the generator and the discriminator optimize the following min-max objective function,

$$\min_\theta \max_\phi \mathcal{L}_{\text{GAN}}(\theta, \phi) \quad (1)$$

$$\mathcal{L}_{\text{GAN}}(\theta, \phi) = \mathbb{E}_{\mathbf{x} \sim p_r(\mathbf{x})}[F(D_\phi(\mathbf{x}))] + \mathbb{E}_{\mathbf{z} \sim p(\mathbf{z})}[F(1 - D_\phi(G_\theta(\mathbf{z})))], \quad (2)$$

where $F : \mathbb{R} \to \mathbb{R}$ is a real-valued function. Note that the generator and discriminator play a min-max game with the generator attempting to minimize the loss $\mathcal{L}_{\text{GAN}}(\theta, \phi)$ while the discriminator attempts to maximize it.

In the standard GAN training, the training procedure employs the log loss objective with $F(x) = \log(\sigma(x))$, where $\sigma(x)$ denotes the *sigmoid* function. Intuitively, this necessitates the discriminator to assign high values to the true sample and low values to the generated, synthetic samples. The resulting $\mathcal{L}_{\text{GAN}}(\theta, \phi)$ can be viewed as minimizing the *Jensen–Shannon divergence* between the real data distribution $p_r$ and the generated data distribution $p_g$. However this loss is susceptible to training instability which significantly contributes to the mode collapse phenomenon where the generator produces a limited variety of outputs [16, 19, 20]. Various loss functions have since been introduced with different choices of $F(x)$ leading to other divergence measures or integral probability metrics. In particular, the choice of $F(x) = x$ was used in the Wasserstein GAN objective function [21]. Wasserstein GANs are known to improve the stability of GANs by mitigating the impact of mode collapse.

Under the assumption that the family of parametrized discriminator functions $\{D_\phi\}_\phi$ are 1-Lipschitz continuous, the min-max optimisation problem for the Wasserstein GAN is defined as [22]

$$\mathcal{L}_{\text{GAN}}(\theta, \phi) = \mathbb{E}_{\mathbf{x} \sim p_r(\mathbf{x})}[D_\phi(\mathbf{x})] - \mathbb{E}_{\mathbf{z} \sim p(\mathbf{z})}[D_\phi(G_\theta(\mathbf{z}))]. \tag{3}$$

In practice, the 1-Lipschitz assumption is enforced by regularizing $\mathcal{L}_{\text{GAN}}(\theta, \phi)$ with a gradient penalty term as [23]

$$\mathcal{L}_{\text{GAN-GP}}(\theta, \phi) = \mathcal{L}_{\text{GAN}}(\theta, \phi) + \lambda \mathbb{E}_{\hat{\mathbf{x}} \sim \hat{p}(\mathbf{x})} \left[ (\|\nabla_{\hat{\mathbf{x}}} D_\phi(\hat{\mathbf{x}})\|_2 - 1)^2 \right]. \tag{4}$$

In (4), $\lambda$ is the penalty coefficient and $\hat{p}(\mathbf{x}) = \epsilon p_r(\mathbf{x}) + (1 - \epsilon) p_g(\mathbf{x})$ corresponds to the distribution of points interpolated between the true distribution $p_r(\mathbf{x})$ and generated distribution $p_g(\mathbf{x})$, where $p_g(\mathbf{x})$ is defined by $\mathbf{x} = G_\theta(\mathbf{z})$ with $\mathbf{z} \sim p(\mathbf{z})$, and $\epsilon$ is sampled from a uniform distribution.

Note that in Wasserstein GANs, the discriminator is renamed as critic. This is because, in contrast to standard GANs where the discriminator assigns binary labels to classify between real and synthetic data, the Wasserstein critic assigns 'scores' to the real and synthetic data samples, without explicitly classifying it. This allows the critic to provide more informative gradients to the generator network during backpropagation, allowing for more stable and meaningful training dynamics in the generator network.

### 2.2 Variational AutoEncoders

Variational AutoEncoder [24] is a latent variable generative model that seeks to generate synthetic data samples based on the principles of variational Bayesian inference. Unlike GANs, which employ an adversarial training mechanism between a generator and a discriminator, VAEs achieve this by maximizing the evidence lower bound (ELBO) of the marginal likelihood of the observed data. Specifically, it optimizes the likelihood of a parametrized distribution $p_\theta(\mathbf{x})$ that approximates the true but unknown data distribution $p_r(\mathbf{x})$ underlying the observed dataset $\mathcal{D} = \{\mathbf{x}_i\}_{i=1}^N$.

A VAE consists of two networks: (a) an *encoder* network, parameterized by $\omega$, that defines the conditional distribution $q_\omega(\mathbf{z}|\mathbf{x})$ of encoding the input data $\mathbf{x} \in \mathcal{X}$ into a lower-dimensional latent representation $\mathbf{z} \in \mathcal{Z}$, and (b) a *decoder* network, parametrized by $\theta$, that defines the conditional distribution $p_\theta(\mathbf{x}|\mathbf{z})$ of decoding the latent vector $\mathbf{z}$ to the data space $\mathcal{X}$. Furthermore, VAE regularizes the encoder by imposing a prior distribution $p(\mathbf{z})$ over the latent space $\mathcal{Z}$. Typical implementations of VAE use a Gaussian prior $p(\mathbf{z}) = \mathcal{N}(0, \mathbb{I})$ with identity covariance matrix, and Gaussian encoder $q_\omega(\mathbf{z}|\mathbf{x}) = \mathcal{N}(\mu_\omega, \sigma_\omega^2 \mathbb{I})$ with $(\mu_\omega, \log \sigma_\omega^2)$ determined by neural networks with parameters $\omega$.

VAE aims to minimize the negative ELBO,

$$\mathcal{L}_{\text{VAE}}(\omega, \theta) = -\mathbb{E}_{\mathbf{x} \sim p_r(\mathbf{x})} \mathbb{E}_{\mathbf{z} \sim q_\omega(\mathbf{z}|\mathbf{x})} \left[ \log \frac{p_\theta(\mathbf{x}|\mathbf{z}) p(\mathbf{z})}{q_\omega(\mathbf{z}|\mathbf{x})} \right] = \mathcal{L}_{\text{recon}}(\omega, \theta) + \mathcal{L}_{\text{prior}}(\omega), \tag{5}$$

where the reconstruction loss $\mathcal{L}_{\text{recon}}(\omega, \theta)$ and the prior regularisation term $\mathcal{L}_{\text{prior}}(\omega)$ are defined as

$$\mathcal{L}_{\text{recon}}(\omega, \theta) = -\mathbb{E}_{\mathbf{x} \sim p_r(\mathbf{x})} \mathbb{E}_{\mathbf{z} \sim q_\omega(\mathbf{z}|\mathbf{x})} [\log p_\theta(\mathbf{x}|\mathbf{z})]$$

and

$$\mathcal{L}_{\text{prior}}(\omega) = \mathbb{E}_{\mathbf{x} \sim p_r(\mathbf{x})} [D_{\text{KL}}(q_\omega(\mathbf{z}|\mathbf{x}) \| p(\mathbf{z}))], \tag{6}$$

with $D_{\text{KL}}(p\|q)$ denoting the Kullback-Leibler divergence between two distributions $p$ and $q$.

### 2.3 Variational Quantum Circuits

Variational quantum circuits, or quantum neural networks, consist of parametrized unitary operators (or quantum gates) applied on an initial $n$-qubit quantum state $|0\rangle^{\otimes n}$. To process classical data samples $\mathbf{z} \in \mathcal{Z}$, they must be first encoded into a quantum state through an appropriate embedding technique. The quantum embedding maps $\mathbf{z} \mapsto |\psi(\mathbf{z})\rangle$ to a quantum state $|\psi(\mathbf{z})\rangle$.

Throughout this work, we use an angle encoding scheme that embeds the classical data $\mathbf{z}$ into the rotation angles of $R_x/R_y/R_z$ gates. Specifically, for $\mathbf{z} \in \mathbb{R}^n$, angle encoding with $R_y$ gate applies a unitary $V(\mathbf{z}) = \prod_{i=1}^n R_y(z_i)$ onto the initial quantum state $|0\rangle^{\otimes n}$ to yield the quantum state

$$|\psi(\mathbf{z})\rangle = V(\mathbf{z}) |0\rangle^{\otimes n}. \tag{7}$$

The embedded quantum state $|\psi(\mathbf{z})\rangle$ is then followed by the application of a parametrized unitary gate $U(\theta)$ on the $n$-qubit system. Conventionally, this unitary is implemented in $L$-layers using $l$-th layer unitary $U_l(\theta_l)$ as $U(\theta) = \prod_{l=1}^L U_l(\theta_l)$. This yields the final parametrized quantum state $|\psi(\mathbf{z}, \theta)\rangle = U(\theta) |\psi(\mathbf{z})\rangle$ determined by the variational quantum circuit. To extract classical information for post-processing, we then apply quantum measurements on the state $|\psi(\mathbf{z}, \theta)\rangle$. The specific measurement scheme we employ will be detailed in Section 3.3.

## 3 VAE-QWGAN Framework

To address the impact of mode collapse in QGANs, we design a hybrid quantum-classical model, termed VAE-QWGAN, that utilizes the latent space representation produced by a VAE encoder as a prior distribution for image generation. In Fig. 1 we illustrate our proposed VAE-QWGAN architecture that makes use of the patch-quantum generator from [10].

As shown in Fig. 1, the VAE-QWGAN combines the QGAN with VAE by collapsing the VAE decoder and QGAN generator into one quantum model with shared parameters $\theta$. Specifically, we use a Gaussian decoder $p_\theta(\mathbf{x}|\mathbf{z}) = \mathcal{N}(G_\theta(\mathbf{z}), \mathbb{I})$ whose mean is determined by the generator function $G_\theta(\mathbf{z})$ modelled by a **variational quantum circuit**. In this work, we adopt a patch-based variational quantum circuit, consisting of multiple sub-circuits as in [10]. We detail the quantum generator architecture used in Section 3.3. However, it is worth noting that the framework easily extends to other quantum models $G_\theta(\mathbf{z})$. In addition to the quantum generator, the VAE-QWGAN consists of an encoder $q_\omega(\mathbf{z}|\mathbf{x})$, defined as in the standard VAE, mapping the true data sample to a point in the latent space, and a discriminator/critic $D_\phi(\cdot)$ that aims to distinguish between true and generated samples.

### 3.1 Training

We train VAE-QWGAN via the combined loss,

$$\mathcal{L}(\theta, \phi, \omega) = \mathcal{L}_{\text{VAE}}(\omega, \theta) + \mathcal{L}_{\text{QGAN-GP}}(\theta, \phi, \omega), \tag{8}$$

where $\mathcal{L}_{\text{VAE}}(\omega, \theta)$ is defined as in (5), and

$$\mathcal{L}_{\text{QGAN-GP}}(\theta, \phi, \omega) = \lambda \mathbb{E}_{\hat{\boldsymbol{x}} \sim \hat{p}(\mathbf{x})} \left[ (\|\nabla_{\hat{\boldsymbol{x}}} D_\phi(\hat{\boldsymbol{x}})\|_2 - 1)^2 \right]$$
$$+ \mathbb{E}_{\mathbf{x} \sim p_r(\mathbf{x})} [D_\phi(\mathbf{x})] - \mathbb{E}_{\mathbf{x} \sim p_r(\mathbf{x})} \mathbb{E}_{\mathbf{z} \sim q_\omega(\mathbf{z}|\mathbf{x})} [D_\phi(G_\theta(\mathbf{z}))] \tag{9}$$

is a modified gradient penalty-based Wasserstein training objective. Importantly, distinct from the conventional QGAN training in (4) that uses latent vectors sampled from prior $p(\mathbf{z})$, our hybrid VAE-QWGAN uses latent vectors sampled from the VAE encoder distribution $q_\omega(\mathbf{z}|\mathbf{x})$ (see (9)). The resulting training loss $\mathcal{L}_{\text{QGAN-GP}}(\theta, \phi, \omega)$ for VAE-QWGAN therefore depends on the parameters of the encoder, the generator and the critic.

One can interpret the training criteria in (8) as a balance between *style* and *content* loss. The content loss stems from the VAE's reconstruction objective $\mathcal{L}_{\text{recon}}(\omega, \theta)$, which encourages each input image $\mathbf{x}$ to be faithfully reconstructed when passed through the encoder and decoder pair. This term is equivalent to the L2 loss and is thus a pixel-level measure of local intensity discrepancies that preserves the essential structure and detail in the reconstructed images. In contrast, the style loss arises from the QGAN feedback $\mathcal{L}_{\text{QGAN-GP}}(\theta, \phi, \omega)$. This term encourages the generated images to align with the real data distribution. By optimizing the Wasserstein distance, the generated samples capture essential distributional features such as shape and characteristic variations in the data. While the content loss enforces pixel-level fidelity via reconstruction, the style loss ensures that the generated outputs remain visually coherent and representative of the variations found in real samples. However, jointly optimizing both objectives can lead to instability during training. To ensure stable convergence of the criteria in (8), we adopt the following practical considerations:

- According to (8), the encoder training depends on the VAE loss as well as the QGAN loss, where the latter's dependence is due to using the encoder distribution $q_\omega(\mathbf{z}|\mathbf{x})$ as the prior. In practice, to prevent the adversarial gradients from backpropagating into the encoder paramaters, we dissociate signals from the QGAN and update the encoder parameters as

$$\omega \xleftarrow{+} -\nabla_\omega \mathcal{L}_{\text{VAE}}(\omega, \theta). \tag{10}$$

**Algorithm 1** VAE-QWGAN: Training
---
**Input:** Gradient penalty coefficient $\lambda$, critic iterations per generator iteration $n_D$, number of epochs $n_{\text{epochs}}$, reconstruction weight $\gamma$, batch size $m$.
**Initialise:** Encoder parameters $\omega$, generator parameters $\theta$, critic parameters $\phi$

1: **for** epoch $= 1, \ldots, n_{\text{epochs}}$ **do**
2:     Randomly sample $\mathbf{X} \subset \mathcal{D} = \{\mathbf{x}_1, \ldots, \mathbf{x}_n\}$ of $m$ data points
3:     $\mathbf{Z} \sim q_\omega(\mathbf{Z}|\mathbf{X})$
4:     $\mathcal{L}_{\text{prior}}(\omega) \leftarrow D_{\text{KL}}(q_\omega(\mathbf{Z}|\mathbf{X}) \| p(\mathbf{Z}))$
5:     $\mathbf{X}' \leftarrow \mathrm{G}_\theta(\mathbf{Z})$
6:     $\mathcal{L}_{\text{recon}}(\omega, \theta) \leftarrow -\log p_\theta(\mathbf{X}|\mathbf{Z})$
7:     // Loop over the number of discriminator updates
8:     **for** t $= 1, \ldots, n_D$ **do**
9:         Sample a random number $\epsilon \sim U[0,1]$
10:         $\hat{\mathbf{X}} \leftarrow \epsilon \mathbf{X} + (1-\epsilon)\mathbf{X}'$
11:         // Compute WGAN loss objective
12:         $\mathcal{L}_{\text{QGAN-GP}}(\theta, \phi, \omega) = D_\phi(\mathbf{X}) - D_\phi(\mathbf{X}') + \lambda \big( \|\nabla_{\hat{\mathbf{X}}} D_\phi(\hat{\mathbf{X}})\|_2 - 1 \big)^2$
13:         // Update critic parameters according to gradients
14:         $\phi^+ \leftarrow -\nabla_\phi \mathcal{L}_{\text{QGAN-GP}}(\theta, \phi, \omega)$
15:     **end for**
16:     // Update encoder and decoder/generator parameters after critic updates
17:     $\omega^+ \leftarrow -\nabla_\omega(\mathcal{L}_{\text{prior}}(\omega) + \mathcal{L}_{\text{recon}}(\omega, \theta))$
18:     $\theta^+ \leftarrow -\nabla_\theta(\gamma \mathcal{L}_{\text{recon}}(\omega, \theta) - \mathcal{L}_{\text{QGAN-GP}}(\theta, \phi, \omega))$
19: **end for**

---

- Balancing style vs content loss: From (8), the generator is trained based on content loss-based signal from the VAE and style loss-based signals from the QGAN. To effectively balance the two losses, following [17], we use a weighing parameter $\gamma > 0$ to balance the contribution of the respective losses to the generator parameter update:

$$\theta \xleftarrow{+} -\nabla_\theta(\gamma \mathcal{L}_{\text{recon}}(\omega, \theta) - \mathcal{L}_{\text{QGAN-GP}}(\theta, \phi, \omega)). \tag{11}$$

The VAE-QWGAN model trained as in Algorithm 1 benefits from using a latent manifold, via the encoder output, that closely aligns with the true data manifold. This approach enhances both the fidelity and diversity of the generated samples, as the generator is less prone to collapsing onto limited modes and instead produces a more representative set of synthetic data.

### 3.2 Inference

In the training outlined above, the generator receives latent vectors sampled from the encoder's posterior distribution $q_\omega(\mathbf{z}|\mathbf{x})$. However, at inference time, we do not have access to input data $\mathbf{x}$, meaning we cannot directly sample from $q_\omega(\mathbf{z}|\mathbf{x})$. Instead, we establish a method for generating latent representations that can be used to synthesize new data at inference time.

To this end, we fit a Gaussian mixture model [25] on the latent vectors $\{\mathbf{z}\}$ generated at the last epoch of the training. Specifically, after the final epoch of training, we collect the latent vectors $\mathbf{z}_i$ corresponding to each training input $\mathbf{x}_i$ and fit a GMM with parameters $(\mu, \Sigma)$ to capture the underlying structure of the latent space. This model then serves as the prior distribution to enable sampling at inference time. During inference, we then generate synthetic samples by sampling latent vectors $\mathbf{z} \sim \text{GMM}(\mu, \Sigma)$ from the learned GMM and then feeding them into the generator $G_\theta(\cdot)$, producing data samples $\hat{\mathbf{x}} = G_\theta(\mathbf{z})$ (see Algorithm 2).

---

**Algorithm 2** Inference for the VAE-QWGAN Model

---

**Input:**
$\mu, \Sigma$: GMM parameters learned from the final epoch's latent embeddings
$\theta$: Trained generator parameters from the final epoch
$N$: Number of samples to generate (optional)

1: **for** $i = 1, \ldots, N$ **do**
2:     Sample a latent vector $\mathbf{z}_i \sim \text{GMM}(\mu, \Sigma)$
3:     Generate a new sample $\hat{\mathbf{x}}_i \leftarrow G_\theta(\mathbf{z}_i)$
4:     **Output** $\hat{\mathbf{x}}_i$
5: **end for**

---

### 3.3 Quantum Generator Architecture

Since the VAE-QWGAN is designed to learn high-dimensional classical datasets, a critical challenge is the design of quantum generator architectures that can efficiently produce high-dimensional data. In light of this, we adopt the patch-based generator architecture from [10], which uses a quantum generator $G_\theta(\mathbf{z}) = [G_{\theta_1}(\mathbf{z}), \ldots, G_{\theta_{N_g}}(\mathbf{z})]^\top$ that concatenates the output of $N_g$ sub-generators. Note that the above patch-based quantum generator architecture operates directly on the high dimensional space of pixels instead of learning a latent space representation of the images in question as in [13, 14].

In this quantum generator architecture, the $j$th quantum sub-generator that returns the vector $G_{\theta_j}(\mathbf{z})$ is implemented via an $n$-qubit **parameterised quantum circuit** (PQC) as shown in Fig. 3. The PQC implements a unitary operator $U_{\theta_j}(\mathbf{z})$ on an initial ground state $|0\rangle^{\otimes n} = |\mathbf{0}\rangle$ to result in the quantum state $|\psi_j(\mathbf{z})\rangle = U_{\theta_j}(\mathbf{z})|\mathbf{0}\rangle$. Specifically, in this paper, we use the unitary operator of the form $U_{\theta_j}(\mathbf{z}) = U(\theta_j)V(\mathbf{z})$ where $V(\mathbf{z}) = \bigotimes_{i=1}^{n} R_y(z_i)$ is an $R_y$-rotation-based angle encoding unitary. The unitary $V(\mathbf{z})$ is followed by $L$ layers of parametrized unitary gates $U(\theta_j) = \prod_{l=1}^{L} W^{(l)} \bigotimes_{i=1}^{n} R(a_{l,i}, b_{l,i}, c_{l,i})$, where $W^{(l)}$ is the CNOT-entangling gate, and $R(a, b, c)$ is the single-qubit general U3 operator

$$R(a, b, c) = \begin{bmatrix} \cos\left(\frac{a}{2}\right) & -e^{ic}\sin\left(\frac{a}{2}\right) \\ e^{ib}\sin\left(\frac{a}{2}\right) & e^{i(b+c)}\cos\left(\frac{a}{2}\right) \end{bmatrix}, \tag{12}$$

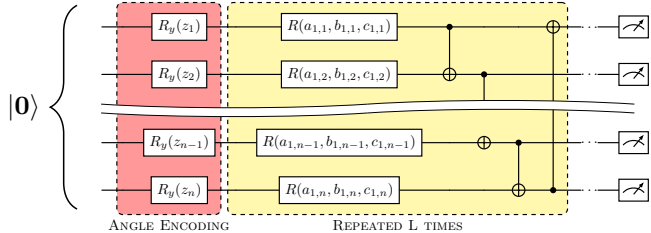

**Fig. 3**: Quantum sub-generator architecture: Hardware efficient ansatz for $n$ qubits with $R_y$ embedding and $L$ layers of repeated U3 rotations and strongly entangling CNOT layers.

defined in terms of parameters $a$, that control rotation around Y-axis, and $b$ and $c$, that introduce phase shifts. The $U3$ gate is the most general single-qubit unitary transformation, capable of performing an arbitrary rotation on the Bloch sphere. The complete set of parameters $\{(a_{l,i}, b_{l,i}, c_{l,i}) : l = 1, \ldots, L, i = 1, \ldots, n\}$ then constitute the vector $\theta_j$. The CNOT-gate $W^{(l)}$ introduces quantum correlations (or entanglement) across qubits, enabling the circuit to generate complex quantum states. We choose a scheme where the entangling operation operates on pair-wise adjacent qubits until the end of the register is reached; here, we reverse the CNOT direction and connect the last qubit to the first.

To obtain the real-valued function $G_{\theta_j}(\mathbf{z})$ from the above-defined PQC, we first perform a non-linear projective measurement as explained below. The quantum register is first split into a set of data qubits $n_d$ and ancillary qubits $n_a$ such that $n_a + n_d = n$. We then apply a projective measurement on the $n_a$ ancillary qubits of the state $|\psi_j(\mathbf{z})\rangle$ of the sub-generator via the observable $O = |0\rangle\langle 0|^{\otimes n_a}$. Following this, we trace out the contribution of the ancillary qubits from the overall state using the partial trace operation $\mathrm{Tr}_{n_a}(\cdot)$. This results in the following $n_d$-qubit mixed state

$$\rho_j(\mathbf{z}) = \mathrm{Tr}_{n_a}\left(\frac{(O \otimes I)|\psi_j(\mathbf{z})\rangle\langle\psi_j(\mathbf{z})|}{\langle\psi_j(\mathbf{z})|(O \otimes I)|\psi_j(\mathbf{z})\rangle}\right), \tag{13}$$

represented as a density matrix. Note that the above approach of a partial measurement on the ancillary qubit system introduces non-linearity in the quantum generator model, which is otherwise not possible with unitary transformations.

We then measure the mixed state $\rho_j(\mathbf{z})$ in the computational basis, to extract classical values that define the function $G_{\theta_j}(\mathbf{z})$. From Born's rule, we have that the probability of measuring the state $\rho_j(\mathbf{z})$ in the computational basis state $|k\rangle$, for $k = 0, \ldots, 2^{n_d} - 1$, is given by $p_j(k) = \mathrm{Tr}(|k\rangle\langle k|\rho_j(\mathbf{z}))$. Through computational basis measurements on $\rho_j(\mathbf{z})$, we then extract the following vector of probabilities,

$$\tilde{\mathbf{x}}^{(j)} = [p_j(0), \ldots, p_j(2^{n_d} - 1)]. \tag{14}$$

However, this probability distribution is not directly suitable for image-based datasets, as pixel values typically require a different range. To transform these outputs into meaningful pixel values, we apply the following additional post-processing step to yield the function $G_{\theta_j}(\mathbf{z})$ as,

$$G_{\theta_j}(\mathbf{z}) = \frac{\tilde{\mathbf{x}}^{(j)}}{\max_k \tilde{\mathbf{x}}_k^{(j)}}\bigg|_{[:P]}, \quad (15)$$

where $\tilde{\mathbf{x}}_k^j$ denotes the $k$-th entry of the vector $\tilde{\mathbf{x}}^{(j)}$, and $P = \frac{HW}{N_g}$ is the number of pixels per patch (with $H$ and $W$ being the height and width of the full image respectively). The slicing operation $\mathbf{a}\big|_{[:P]}$ selects the first $P$ components of the vector $\mathbf{a}$. This ensures that the highest probability maps to intensity 1, while preserving relative relationships among the other values. Finally, we construct the output image $G(\mathbf{z})$ by concatenating the outputs from all $N_g$ sub-generators:

$$G(\mathbf{z}) = \text{concat}\left(G_{\theta_1}(\mathbf{z}), \ldots, G_{\theta_{N_g}}(\mathbf{z})\right), \quad (16)$$

with each sub-generator contributing a portion of the final output before being pieced together based on a specific patch layout strategy.

### 3.4 Encoder and Critic

The encoder network is designed as a deep convolutional architecture that progressively reduces the spatial dimensions of the input image to extract hierarchical features. In the implementation, three sequential convolutional layers are employed with kernel sizes of 4, strides of 2, and padding of 1 to ensure a smooth reduction in resolution while preserving image characteristics. Each convolutional layer is followed by a LeakyReLU activation function to mitigate the vanishing gradient problem by allowing a small gradient when inputs are negative [26]. After these layers, the multidimensional output is flattened into a one-dimensional vector where this flattened representation is then fed into a fully connected layer to form an intermediate high-level feature space. From this space, two distinct linear transformations compute the mean and logarithmic variance parameters necessary for the reparameterization trick in variational autoencoding. We use the Kaiming normal initialization strategy to maintain a stable variance across layers and support efficient training [27].

The critic network is implemented as a dense neural network structured to distinguish real images from those generated by the model. Unlike the encoder, the critic begins by flattening the input image into a vector that is subsequently processed through a series of three fully connected layers. The first two layers map the input to 512 and then 256 neurons, with each layer followed by a LeakyReLU activation function. The final layer outputs a single scalar value that quantifies the critic's confidence in the authenticity of the input. This dense architecture is purposefully designed to deliver a strong and stable learning signal to the generator during adversarial training. Consistent with the encoder, the critic's weights are also initialized using the Kaiming normal initialization method.

## 4 Experimental Settings

The experimental implementation is carried out in Python 3.11.4 using PyTorch [28] and PyTorch Lightning [29]. PyTorch is an open-source machine learning framework that provides a flexible platform for hybrid neural network execution. We use PennyLane [30] for quantum circuit construction and optimization, leveraging its seamless integration with PyTorch for automatic differentiation and hybrid quantum-classical machine learning workflows.

### 4.1 Evaluation Metrics

Unlike classification tasks where the final evaluation metric of accuracy is relatively straightforward, evaluating the performance of a generative model is more complex and often requires multiple metrics to capture different aspects of the model's performance. We use several different metrics across our experiments to assess both the sample quality and mode-capturing phenomenon in GANs that include Fréchet Distance (FD), Jensen Shannon Divergence (JSD), Number of statistically Distinct Bins (NDB), Peak Signal-To-Noise Ratio (PSNR), Structural Similarity Index Measure (SSIM) and cosine similarity.

Fréchet Distance[1] (FD) evaluates the quality of generated images by measuring the statistical similarity between real and generated data distributions within a feature space defined by flattened raw pixel intensities. Specifically, each image of size $28 \times 28$ is represented as a 784-dimensional vector, allowing the metric to capture direct pixel-level variations. Under the assumption that both real data distribution $p_r(\mathbf{x})$ and generated data distribution $p_g(\mathbf{x})$ follow multivariate Gaussian distributions, FD is mathematically defined as

$$\mathrm{FD}(p_r(\mathbf{x}), p_g(\mathbf{x})) = |\mu_r - \mu_g|^2 + \mathrm{Tr}\left(\Sigma_r + \Sigma_g - 2\sqrt{\Sigma_r \Sigma_g}\right) \quad (17)$$

where $\mu_r$ and $\mu_g$ denote the means of the real and generated distributions respectively, and $\Sigma_r$ and $\Sigma_g$ represent their corresponding covariance matrices. The mean and covariance matrices are estimated from the true and generated data samples, and then plugged in (17) to compute the FD [13, 33]. A lower value of FD indicates greater similarity between the generated samples and the real data, signifying improved quality and fidelity of the generated images.

The Jensen Shannon Divergence (JSD) and Number of statistically Different Bins (NDB) are used to evaluate the diversity of the generated distribution, thereby providing a measure for mode collapse [34, 35]. Both the measures are better suited to evaluate discrete distributions. In our experiments with continuous datasets, we first use the K-means algorithm to cluster the training data samples into $K$ clusters (or bins). The resulting histogram describes a discretized distribution $Q$ of the true underlying data distribution $r$. We set $Q$ as the target distribution.

---

[1] Fréchet Inception Distance (FID), commonly used for image quality evaluation, is inappropriate in this context since the Inception-V3 network [31] utilized for computing FID is trained on 3-channel images of size $299 \times 299$. Our experiments involve grayscale images of size $28 \times 28$, making direct pixel-based Fréchet Distance a more suitable metric [32].

For discretizing the generated distribution, we assign each of the generated data samples to the closest cluster centroid, based on the Euclidean $L_2$ distance. The resulting histogram describes the generated discrete distribution $P$. The JSD between $P$ and $Q$ can be then computed as

$$D_{\text{JS}}(P\|Q) = \frac{1}{2}D_{\text{KL}}\left(P\left\|\frac{1}{2}(P+Q)\right.\right) + \frac{1}{2}D_{\text{KL}}\left(Q\left\|\frac{1}{2}(P+Q)\right.\right).$$

In contrast to the commonly used Kullback-Leibler divergence, the JSD is symmetric and is bounded between 0 and 1, making it a more stable and interpretable metric for comparing probability distributions. A lower value of JSD indicates greater similarity between the generated and target distributions, implying better diversity and reduced mode collapse in the generated data [14, 34].

To calculate the NDB measure, we first evaluate the proportion of real samples and generated samples that fall into each bin. A two-sample test (such as a z-test) is then applied to check whether the proportions $Q$ and $P$ are significantly different. The bins where the proportions of real and generated samples are statistically different are counted as distinct. The NDB is the total number of such distinct bins, providing a measure of how well the generated samples cover the different modes (bins) of the real data distribution. In this work, we report the NDB/K, the number of statistically different bins normalised by the number of bins.

Peak Signal to Noise Ratio (PSNR), Structural Similarity Index Measure (SSIM), and Cosine Similarity assess similarity between generated and real images from different perspectives. The PSNR measures pixel-wise similarity by comparing the signal power of the original images against the noise power introduced by reconstruction [36]

$$\text{PSNR} = 10\log_{10}\left(\frac{\text{MAX}^2}{\text{MSE}}\right), \quad (18)$$

where MAX is the maximum possible pixel value, and MSE is the mean squared error between real and generated images. A higher PSNR value indicates less distortion and higher image fidelity.

Structural Similarity Index Measure (SSIM) evaluates perceptual similarity by comparing luminance, contrast, and structural information between two images [36]. Given a real image $\mathbf{x}$ and a generated image $\mathbf{x}'$, SSIM is defined as:

$$\text{SSIM}(\mathbf{x},\mathbf{x}') = \frac{(2\mu_x\mu_{x'} + c_1)(2\sigma_{xx'} + c_2)}{(\mu_x^2 + \mu_{x'}^2 + c_1)(\sigma_x^2 + \sigma_{x'}^2 + c_2)}, \quad (19)$$

where $\mu_x$ and $\mu_{x'}$ represent the pixel-wise average intensity (mean luminance) of the true and generated images, $\sigma_x^2$ and $\sigma_{x'}^2$ denote the pixel intensity variance (contrast) of the true and generated images, and $\sigma_{xx'}$ denotes the covariance between the two images' pixel intensities. SSIM yields a scalar similarity score between -1 and 1, with values closer to 1 indicating higher structural similarity.

Cosine similarity quantifies the semantic consistency between real and generated images by measuring the angle between their respective flattened pixel representations. Each real ($\mathbf{x}$) and generated ($\mathbf{x}'$) image of size $28 \times 28$ is reshaped into a 784-dimensional vector of pixel intensities. Cosine Similarity is then defined as:

$$\text{Cosine Similarity}(\mathbf{x}, \mathbf{x}') = \frac{\mathbf{x} \cdot \mathbf{x}'}{\|\mathbf{x}\|\|\mathbf{x}'\|}, \tag{20}$$

where $\|\mathbf{x}\|$ and $\|\mathbf{x}'\|$ denote the Euclidean norms of the real and generated image vectors, respectively. The resulting value ranges between $-1$ and $1$, with values closer to 1 indicating greater alignment in pixel-space orientation and thus higher semantic similarity. In our implementation, the resulting cosine similarity score is scaled linearly to a range of $[0, 1]$ using the transformation $rescaled = 0.5 + 0.5 \times original$ to ensure a more intuitive interpretation of the similarity measure, where higher scores consistently represent better similarity.

### 4.2 Network Hyperparamaters

For the experiments, we use the MNIST [37] and Fashion-MNIST [38] datasets ($28 \times 28 \times 1$ pixels). We use a quantum generator with $N_g = 14$ sub-generators (see Fig. 3), each consisting of $L = 12$ layers generating patches of shape $(2, 28)$. Each sub-generator has $n = 7$ qubits in total, including one ancilla qubit used for the non-linear partial measurement, yielding a total of 3528 parameters for the entire generator. The weights of each sub-generator are randomly initialized from the uniform distribution $U_{[0,2\pi]}$ and we evaluate (14) in the infinite shot limit.

For parameter optimization of the VAE-QWGAN, we employ the Adam optimizer [39] with a learning rate $lr = 0.01$ for the decoder/generator, $lr = 0.0003$ for the classical encoder and $lr = 0.0005$ for the critic, with the 1st and 2nd momentum terms set as $\beta_1 = 0$ and $\beta_2 = 0.9$ for all optimizers. We set the style vs content loss weighing parameter to be $\gamma = 0.0005$ following [17]. Furthermore, the gradient penalty coefficient is set as $\lambda = 10$. These hyperparameters are fixed based on repeated empirical evaluations to yield the best convergence and stability of the model. For training, we optimise our model for $n_{epochs} = 15$ epochs, with the encoder/decoder parameters updated after every $n_{critic} = 5$ critic parameter updates.

**Inference algorithm ablation study:** We conduct an ablation study on our GMM inference algorithm to evaluate various GMM configurations and identify the optimal model for capturing the latent vector distribution from our VAE-based approach. We explored a grid of configurations, varying both the number of components and covariance types. Specifically, the number of components considered was in the range $[1, 7]$, accommodating different multimodality levels in the latent space. Covariance structures tested included spherical, tied, diagonal, and full, enabling assessment of how covariance flexibility affects model performance. Each GMM configuration was evaluated using the Bayesian Information Criterion (BIC) [40], calculated as $\text{BIC} = -2 \log L + p \log n$, where $L$ denotes likelihood, $p$ parameter count, and $n$ data points. BIC balances model complexity and fit by penalizing overly complex models and favouring better likelihood. The configuration with the lowest BIC was selected as optimal. We performed our ablation study on the models training on the MNIST data and reused the same GMM hyperparameters for the Fashion MNIST model. For the binary class MNIST dataset, the best number of GMM components was 2 using full covariance type with a BIC score of 6170.11. Additionally, for the triple-class dataset, the best number of GMM components was 3, again with a full covariance matrix, here we compute a BIC score of 5849.59. In Fig. 4 we plot a bar chart for our hyperparameter ablation study.

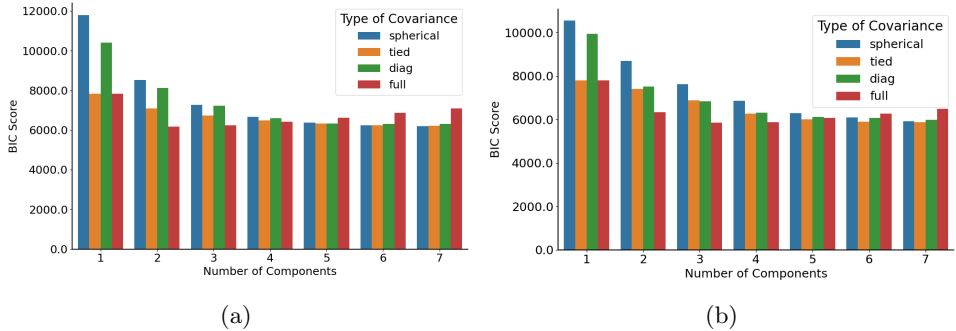

**Fig. 4**: BIC scores for different GMM configurations. Lower BIC scores indicate better model fit. **(a)** BIC scores for the binary-class MNIST, where the optimal configuration is a 2-component GMM with full covariance (BIC = 6170.11). **(b)** BIC scores for the triple-class MNIST dataset, where the optimal model utilizes 3 components with full covariance (BIC = 5849.59).

## 5 Results and Discussion

In this section, we present our main results that compare the performance of our proposed VAE-QWGAN model on the MNIST and Fashion MNIST datasets, against benchmark classical and quantum generative models.

### 5.1 Results on MNIST dataset

**Binary MNIST**: We first consider the generated images using a binary sub-sampling of the MNIST dataset. Specifically, we select two classes – digits '0' and '1' – to generate simultaneously from this dataset. We randomly select 2600 samples from the datasets to serve as our training samples. For binary image generation, we use a mini-batch size $m = 8$, resulting in 325 iterations per epoch of training.

Figure 5a plots the Wasserstein distance between the true and generated data distributions during the training of the VAE-QWGAN as a function of the training epochs. Note that a lower distance indicates a better approximation of the real data distribution. We compare the training performance of VAE-QWGAN with that of the PQWGAN [10] that uses standard normal prior distribution (PQWGAN + $\mathcal{N}(0, \mathbb{I})$) and uniform prior distribution (PQWGAN + $U_{[0,1)}$), and with that of classical GANs under a uniform prior distribution as used in [10]. Figure 5a shows that VAE-QWGAN consistently achieves a lower Wasserstein distance compared to the PQWGAN models, indicating better convergence towards the target distribution.

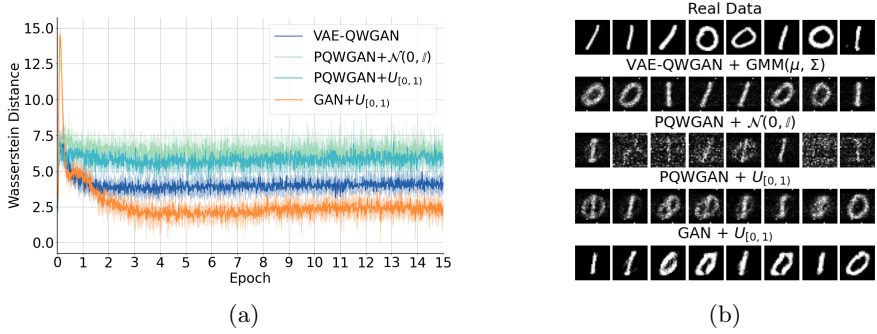

**Fig. 5**: Plots for Binary MNIST: **(a)** Wasserstein distance as a function of training epochs; **(b)** Comparison of images generated by different models.

In Fig. 5b, we visualise and compare the images generated by the trained models. We generate samples from the models PQWGAN + $\mathcal{N}(0, \mathbb{I})$, PQWGAN+$U_{[0,1)}$, the classical GAN with uniform prior $U_{[0,1)}$, and from our proposed VAE-QWGAN +GMM($\mu, \Sigma$) that uses GMM-based inference. It can be seen that VAE-QWGAN better generates the two classes with a clearer separation between the modes. This can be attributed to achieving a lower Wasserstein distance during the training (see Fig. 5a). In contrast, the PQWGAN model with a Gaussian prior achieves the worst Wasserstein distance during training. The resulting generated samples demonstrate a high degree of mode collapse where the images are no longer sharp and there is a clear overlap between the two classes. While these features are also present in the PQWGAN model with a uniform prior, the generated images are sharper. Overall the quantum models lack some sharpness which we attribute to the patch-based generator mechanism used to produce synthetic samples, which lacks flexibility in what pixel values can be taken for image generation. We conjecture that replacing patch-quantum GAN with other efficient quantum generator architectures in the VAE-QWGAN framework can mitigate this issue.

Finally, we note that the classical model incurs the least Wasserstein distance and generates sharper images. This is due to the improved expressivity of these GANs that use significantly higher number of parameters. For the classical GAN model employed here, the generator is composed of almost 1.46 million trainable parameters. In contrast, the QWGAN model—used as the backbone for both our hybrid architecture and the quantum benchmark—requires only 3528 trainable parameters in its patch-based quantum generator. Additionally, the encoder network in the VAE-QWGAN contributes 313 thousand parameters to autoencode the prior distribution. Despite the significantly reduced parameter count, the hybrid quantum models achieve competitive performance in comparison to the classical model, underscoring the expressive capacity of quantum circuits [41]. Notably, a classical generator with a similar number of parameters fails to learn a meaningful data distribution.

To better understand the insights from Fig. 5, we visualise the embedding space of the generated images using t-SNE projections in Fig. 6 [42]. This gives us a clear depiction of how well the PQWGAN with different priors can capture the underlying structure of the data. The t-SNE projections allow us to observe the clustering and separation of the generated samples in a reduced two-dimensional space, providing insight into the extent of mode collapse and class separation.

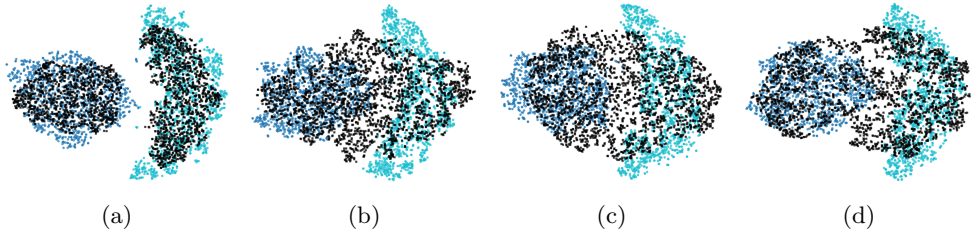

(a)      (b)      (c)      (d)

**Fig. 6**: t-SNE visualizations of the embedding space for Binary MNIST data for **(a)** VAE-QWGAN + GMM$(\mu, \Sigma)$, **(b)** PQWGAN + $\mathcal{N}(0, \mathbb{I})$, **(c)** PQWGAN + $U_{[0,1)}$ and **(d)** classical GAN + $U_{[0,1)}$. The blue and turquoise points represent the t-SNE projections of real data samples corresponding to digits **0** and **1**, respectively. The black points denote the projections of generated (fake) samples from the trained generative models, mapped into the same space as the real images.

From Fig. 6, we see that the latent space embeddings generated by VAE-QWGAN (Fig. 6a) exhibit well-defined clusters, capturing the distinct modes of the dataset. The clusters also closely resemble the real data embedding space. In contrast, embeddings from the uniform prior in Fig. 6c and the Gaussian prior in Fig. 6b show less distinct and more scattered projections. This is due to the use of un-informed prior distributions which do not necessarily sample from a meaningful region of the latent space. Fig. 6d shows the projections for the classical GAN model, demonstrating better coverage over the distinct modes with some interpolation between the two modes.

In Table 1 below, we report the performance of our VAE-QWGAN compared to the other quantum benchmark models. Our evaluation is done on the evaluation metrics detailed in Section 4.1. As we show, our VAE-QWGAN overwhelmingly demonstrates superior image quality and diversity across multiple metrics compared to the other priors.

In Fig. 7, we track the above validation metrics at the end of each training epoch for our quantum benchmark models. For our VAE-QWGAN model, at the end of each training epoch, we use 2600 test images from binary MNIST as input to VAE-QWGAN. We then consider the corresponding generated images, which reconstruct the test inputs, to evaluate the validation metrics. We note that reconstruction of images is not possible in the conventional PQWGAN due to the absence of the encoder network. Therefore, at the end of each training epoch, we sample 2600 latent vectors to generate fake image samples and compute the metrics on these images. In the figures, we evaluate each model according to the JSD, NDB/K, SSIM, Cosine Similarity, PSNR and Fréchet distance.

| Metric (Dataset) | PQWGAN + $\mathcal{N}(0,\mathbb{I})$ | PQWGAN + $U_{[0,1)}$ | VAE-QWGAN + GMM$(\mu,\Sigma)$ |
|---|---|---|---|
| JSD ↓ (BMNIST) | $0.284 \pm 0.0351$ | $0.186 \pm 0.0357$ | $\mathbf{0.126 \pm 0.0175}$ |
| NDB/K ↓ (BMNIST) | $0.967 \pm 0.0408$ | $0.880 \pm 0.0767$ | $\mathbf{0.760 \pm 0.0551}$ |
| SSIM ↑ (BMNIST) | $0.0859 \pm 0.000827$ | $\mathbf{0.148 \pm 0.00584}$ | $0.141 \pm 0.00229$ |
| PSNR ↑ (BMNIST) | $10.3 \pm 0.0395$ | $10.8 \pm 0.134$ | $\mathbf{10.9 \pm 0.0855}$ |
| Cosine Similarity ↑ (BMNIST) | $0.714 \pm 0.00143$ | $\mathbf{0.755 \pm 0.00377}$ | $0.745 \pm 0.00155$ |
| FD ↓ (BMNIST) | $40.6 \pm 0.474$ | $31.8 \pm 1.12$ | $\mathbf{22.6 \pm 1.20}$ |

**Table 1**: Evaluation metrics for images generated by PQWGAN with Uniform and Gaussian priors, and VAE-QWGAN with GMM inference after training on Binary MNIST (BMNIST). Scores are reported as mean ± standard deviation from 5 repeated experiments, 3 s.f reported. ↑ indicates higher values are better, while ↓ indicates lower values are preferable.

Specifically, Fig. 7a and Fig. 7b show how the JSD and NDB/K metrics vary with the different benchmark quantum models. Our VAE-QWGAN achieves a 30.1% reduction in JSD and 7.73% reduction in NDB/K compared to the next best model - the PQWGAN+$U_{[0,1)}$. The consistently lower JSD and NDB/K scores for VAE-QWGAN indicate greater image diversity and reduced mode collapse compared to PQWGAN across different priors, highlighting the effectiveness of VAE-QWGAN in capturing the data distribution accurately. Finally, we observe that the SSIM and PSNR scores on the reconstructed images of VAE-QWGAN improve slightly than that evaluated on the GMM inference based generated samples in Table 1. For PSNR, this can, for instance, be attributed to the mean-squared error computation with respect to true data samples, which for reconstructed images is expected to be lower than that for a synthetically generated image.

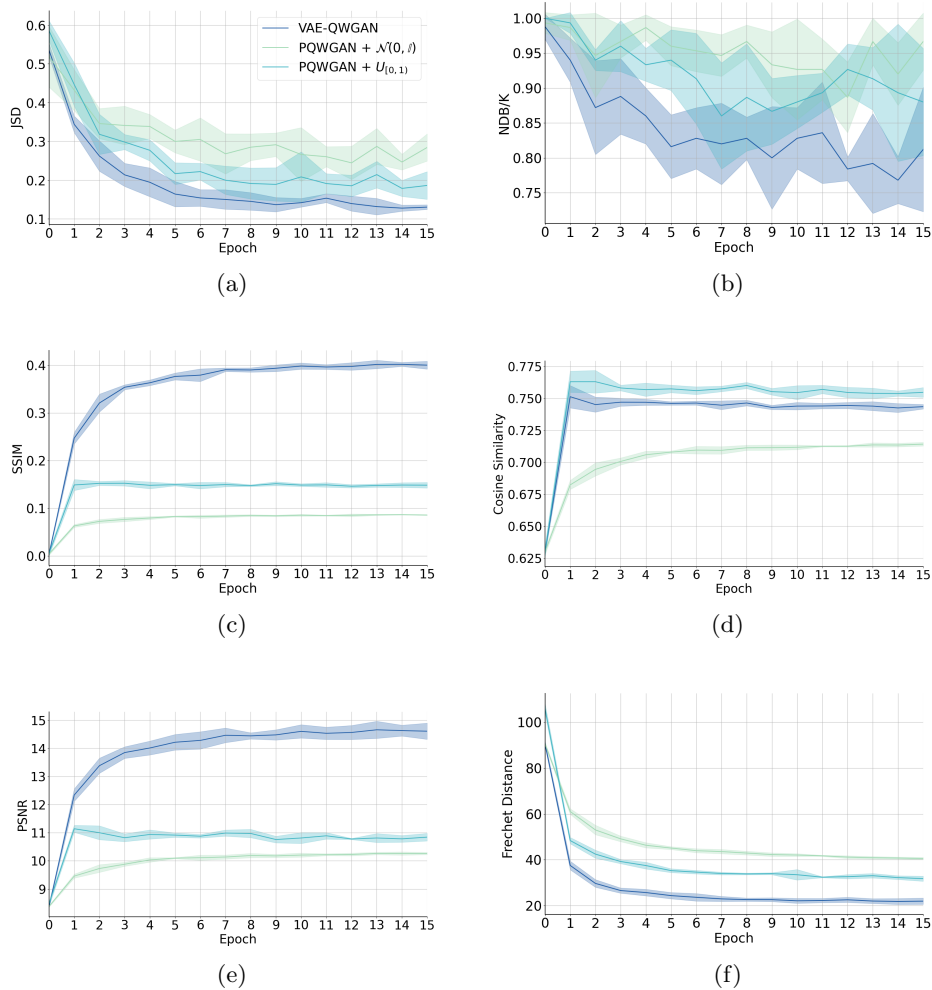

**Fig. 7**: Plots for Binary MNIST evaluating various metrics – **(a)** JSD **(b)** NDB/K **(c)** SSIM **(d)** Cosine Similarity **(e)** PSNR **(f)** FD – as a function of the training epochs for different benchmark quantum models.

**Triple MNIST**: We next evaluate our proposed VAE-QWGAN model using the Triple MNIST dataset consisting of digits '0', '1', and '7'. Like our previous experiments on binary MNIST, we randomly select 2600 samples from these three classes to serve as training data; however, we now set a mini-batch size of $m = 16$. In Fig. 8a, we compare the Wasserstein distance between the real and generated distributions during the training phase of various models on triple MNIST dataset. Our VAE-QWGAN consistently achieves lower Wasserstein distance throughout the training, clearly outperforming other quantum-based models on 3-class image generation. This increased performance is evident from the high quality of corresponding generated images in Fig. 8b. Images from our model are visibly more distinct than the other quantum approaches, with the PQWGAN + $U_{[0,1)}$ prior having more blurry samples and the PQWGAN + $\mathcal{N}(0,\mathbb{I})$ prior failing to even moderately display the different modes. The classical GAN achieves the lowest Wasserstein distance resulting in sharper images. However, as we observe from the generated samples, this doesn't necessarily translate to image quality, as the images have many artifacts and gaps which make distinguishing digits more difficult.

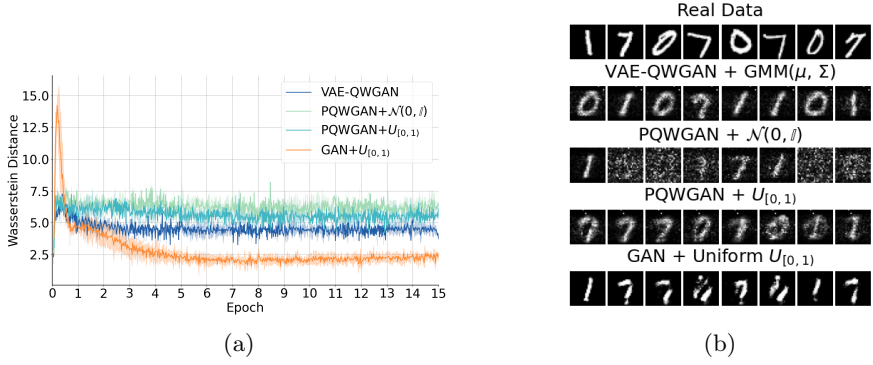

(a)                          (b)

**Fig. 8**: Plots for Triple MNIST: **(a)** Wasserstein distance as a function of training epochs; **(b)** Comparison of images generated by different models.

To understand these observations again we visualise the embedding space of the generated images using t-SNE projections in Fig. 9. In the t-SNE projections, we observe the strong clustering and separation of our VAE-QWGAN even for three class image generation.

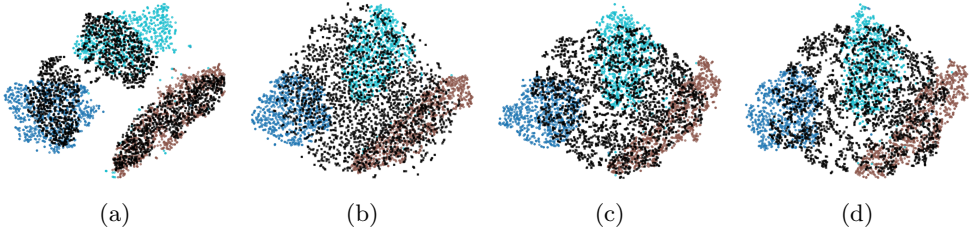

(a)         (b)         (c)         (d)

**Fig. 9**: t-SNE visualizations of the embedding space for Triple MNIST data for **(a)** VAE-QWGAN + GMM($\mu, \Sigma$), **(b)** PQWGAN + $\mathcal{N}(0,\mathbb{I})$, **(c)** PQWGAN + $U_{[0,1)}$ and **(d)** classical GAN + $U_{[0,1)}$. The blue, brown and turquoise points represent the t-SNE projections of real data samples corresponding to digits **0**, **1**, and **7**, respectively. The black crosses denote the projections of fake samples from the trained models.

In Table 2 below, we report the performance of our VAE-QWGAN compared to the benchmark quantum models – PQWGAN + $\mathcal{N}(0, \mathbb{I})$, PQWGAN+$U_{[0,1)}$ – on the evaluation metrics detailed in Section 4.1. We see that our results carry across and the VAE-QWGAN consistently achieves significantly better scores on most of the metrics.

| Metric (Dataset) | PQWGAN + $\mathcal{N}(0, \mathbb{I})$ | PQWGAN + $U_{[0,1)}$ | VAE-QWGAN + GMM$(\mu, \Sigma)$ |
|---|---|---|---|
| JSD ↓ (TMNIST) | $0.387 \pm 0.0544$ | $0.261 \pm 0.00975$ | $\mathbf{0.148 \pm 0.00725}$ |
| NDB/K ↓ (TMNIST) | $0.927 \pm 0.0231$ | $0.880 \pm 0.0529$ | $\mathbf{0.773 \pm 0.0838}$ |
| SSIM ↑ (TMNIST) | $0.0750 \pm 0.00277$ | $\mathbf{0.130 \pm 0.00184}$ | $0.124 \pm 0.00357$ |
| PSNR ↑ (TMNIST) | $10.1 \pm 0.0570$ | $10.8 \pm 0.0857$ | $\mathbf{10.9 \pm 0.0207}$ |
| Cosine Similarity ↑ (TMNIST) | $0.706 \pm 0.00550$ | $\mathbf{0.750 \pm 0.000801}$ | $0.738 \pm 0.0011$ |
| FD ↓ (TMNIST) | $38.9 \pm 0.568$ | $31.5 \pm 0.517$ | $\mathbf{25.8 \pm 0.440}$ |

**Table 2**: Evaluation metrics for images generated by PQWGAN with Uniform and Gaussian priors, and VAE-QWGAN with GMM inference after training on Triple MNIST (TMNIST). Scores are reported as mean ± standard deviation from 3 repeated experiments, 3 s.f. reported. ↑ indicates higher values are better, while ↓ indicates lower values are preferable.

*Impact of Transfer Learning:* To enhance convergence speed and potentially improve the diversity of the generated images, we investigate a transfer learning (TL)-inspired parameter initialization scheme for training on the Triple MNIST dataset. This approach involves initializing our VAE-QWGAN model parameters with the converged model parameters obtained from the Binary MNIST training. Such an initialization scheme capitalizes on previously captured digit-specific features and latent representations, thus guiding the model more rapidly towards convergence and mitigating mode collapse. We perform similar experiments as in the previous sub-sections and report them in Appendix 6 in Fig. 12. We find that VAE-QWGAN initialized via transfer learning consistently outperforms the standard VAE-QWGAN across most metrics. We observe a reduction in the JSD and the NDB/K metrics, indicating an improved mitigation of mode collapse. Furthermore, the transfer-learned model converges faster than standard VAE-QWGAN.

### 5.2 Fashion MNIST dataset results

We now evaluate our proposed VAE-QWGAN model against the Fashion MNIST dataset. Specifically, we consider three distinct categories - 'T-Shirt/Top', 'Trouser', and 'Sneaker' to examine the capability of our VAE-QWGAN and compare to the benchmark quantum models.

Like the Triple MNIST experiments, we randomly select 2600 samples across these classes as our training dataset, maintaining a mini-batch size of $m = 16$. In Fig. 10a, we compare the training dynamics of the different models by comparing the Wasserstein distance, while in Fig. 10b we provide samples of generated images from the different models. In line with our previous results, the performance of our VAE-QWGAN model is superior to the other quantum benchmarks, achieving a lower Wasserstein distance and producing more distinguishable images with fewer artifacts compared to other priors. Interestingly, while the classical GAN model attains the lowest Wasserstein distance, its generated images still exhibit noticeable artifacts, highlighting the additional challenge of the Fashion MNIST dataset.

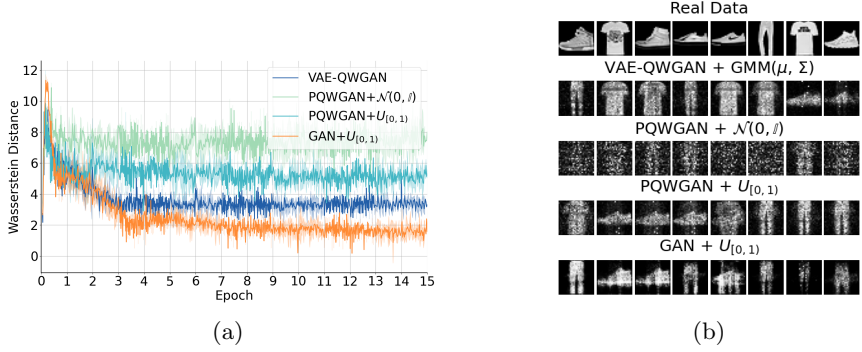

**Fig. 10**: Plots for Triple Fashion MNIST: **(a)** Wasserstein distance as a function of training epochs; **(b)** Comparison of images generated by different models.

We can see from Fig. 11 the embedding space structure visualized via t-SNE projections. These results illustrate again the ability of our model to produce defined clusters which improves the diversity and image quality.

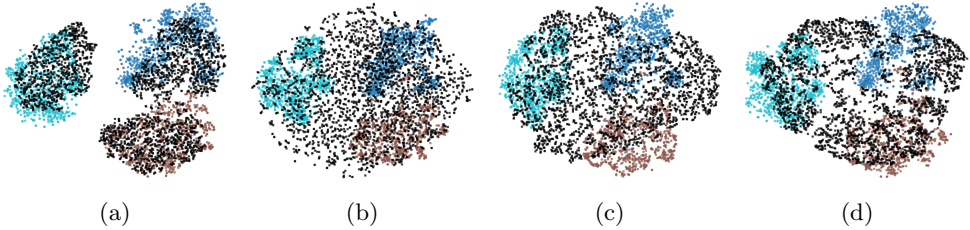

**Fig. 11**: t-SNE visualizations of the embedding space for Triple Fashion-MNIST data for **(a)** VAE-QWGAN + GMM$(\mu, \Sigma)$, **(b)** PQWGAN + $\mathcal{N}(0, \mathbb{I})$, **(c)** PQWGAN + $U_{[0,1)}$ and **(d)** classical GAN + $U_{[0,1)}$. The blue, brown and turquoise points represent the t-SNE projections of real data samples corresponding to labels 'T-shirt', 'Trouser', and 'Sneaker', respectively. The black crosses denote the projections of generated (fake) samples from the trained generative models, mapped into the same space as the real images.

In Table 3 below, we report the performance of our VAE-QWGAN compared to the benchmark quantum models – PQWGAN + $\mathcal{N}(0, \mathbb{I})$, PQWGAN + $U_{[0,1)}$ – on the evaluation metrics detailed in Section 4.1. Our model achieves better scores in diversity metrics – JSD and NDB/K – and in Fréchet distance. However, the PQWGAN +

$\mathcal{N}(0, \mathbb{I})$ model can be seen to achieve better SSIM, PSNR and cosine similarity scores. This suggests that the samples from this model are more visually similar to the real data but are less representative of the broader distribution of real samples.

| Metric (Dataset) | PQWGAN + $\mathcal{N}(0, \mathbb{I})$ | PQWGAN + $U_{[0,1)}$ | VAE-QWGAN + GMM$(\mu, \Sigma)$ |
|---|---|---|---|
| JSD ↓ (TFMNIST) | $0.461 \pm 0.00956$ | $0.326 \pm 0.00603$ | $\mathbf{0.274 \pm 0.0212}$ |
| NDB/K ↓ (TFMNIST) | $0.973 \pm 0.046$ | $0.893 \pm 0.0306$ | $\mathbf{0.88 \pm 0.0163}$ |
| SSIM ↑ (TFMNIST) | $0.0823 \pm 0.00143$ | $\mathbf{0.159 \pm 0.00610}$ | $0.147 \pm 0.00495$ |
| PSNR ↑ (TFMNIST) | $9.21 \pm 0.00572$ | $\mathbf{10.2 \pm 0.137}$ | $9.98 \pm 0.0406$ |
| Cosine Similarity ↑ (TFMNIST) | $0.775 \pm 0.00129$ | $\mathbf{0.837 \pm 0.00201}$ | $0.812 \pm 0.0052$ |
| FD ↓ (TFMNIST) | $57.3 \pm 0.706$ | $33.6 \pm 2.33$ | $\mathbf{25.1 \pm 1.51}$ |

**Table 3**: Evaluation metrics for images generated by PQWGAN with Uniform and Gaussian priors, and VAE-QWGAN with GMM inference after training on Triple Fashion MNIST (TFMNIST). Scores are reported as mean ± standard deviation from 3 repeated experiments, 3 s.f. reported. ↑ indicates higher values are better, while ↓ indicates lower values are preferable.

Finally, in Appendix 6, we investigate the impact of transfer learning with initial weights of our VAE-QWGAN model transferred from the model trained using Binary Fashion MNIST. See Fig. 13 to see the performance of the TL-based VAE-QWGAN across various metrics.

## 6 Conclusion

In this work, we present VAE-QWGAN, a hybrid quantum-classical generative framework that combines the representational power of a classical VAE with the generative capabilities of a QWGAN. By aligning the prior distribution of the QWGAN with the latent space of a trained VAE encoder, the model benefits from a semantically meaningful and data-aware latent manifold that mitigates mode collapse and guides the quantum generator towards high-quality outputs. Furthermore, our inference-time sampling strategy—built on a GMM fitted to the VAE latent space—ensures diverse, high quality image generation. Our empirical results on the MNIST and Fashion-MNIST datasets show clear improvements in both fidelity and diversity of generated samples over state-of-the-art pixel-based hybrid quantum GAN models.

Moving forward, we identify two key directions for advancing this research. Firstly, it is important to investigate qubit-efficient encoding strategies for loading classical data into quantum registers, as well as highly expressive quantum generator architectures. The choice of the encoding method and generator architecture affect the model expressivity and convergence, particularly when the target data distribution is highly complex and high dimensional. Additionally, we will investigate different choices of prior distributions in the VAE objective than the standard normal prior used in this work. This include diffusion-based priors or normalizing flows that can better structure the latent space and navigate the generative process. Adapting our VAE-QWGAN framework with different prior choices, encoding methods, and quantum architectures could potentially further improve mode collapse and generate diverse, high-quality images.

**Funding** No funding source to declare.

# References

[1] John Preskill. "Quantum Computing in the NISQ era and beyond". In: *Quantum* 2 (Aug. 2018), p. 79. ISSN: 2521-327X. DOI: 10.22331/q-2018-08-06-79. URL: https://doi.org/10.22331/q-2018-08-06-79.

[2] Jacob Biamonte et al. "Quantum machine learning". In: *Nature* 549.7671 (2017), pp. 195–202. DOI: https://doi.org/10.1038/nature23474.

[3] Jinkai Tian et al. "Recent Advances for Quantum Neural Networks in Generative Learning". In: *IEEE Transactions on Pattern Analysis and Machine Intelligence* 45.10 (2023), pp. 12321–12340. DOI: 10.1109/TPAMI.2023.3272029.

[4] Christa Zoufal, Aurélien Lucchi, and Stefan Woerner. "Quantum generative adversarial networks for learning and loading random distributions". In: *npj Quantum Information* 5.1 (2019), p. 103.

[5] Alice Barthe et al. "Expressivity of parameterized quantum circuits for generative modeling of continuous multivariate distributions". In: *arXiv preprint arXiv:2402.09848* (2024).

[6] Carlos Bravo-Prieto et al. "Style-based quantum generative adversarial networks for Monte Carlo events". In: *Quantum* 6 (Aug. 2022), p. 777. ISSN: 2521-327X. DOI: 10.22331/q-2022-08-17-777. URL: https://doi.org/10.22331/q-2022-08-17-777.

[7] Ian Goodfellow et al. "Generative adversarial networks". In: *Commun. ACM* 63.11 (Oct. 2020), pp. 139–144. ISSN: 0001-0782. DOI: 10.1145/3422622. URL: https://doi.org/10.1145/3422622.

[8] Seth Lloyd and Christian Weedbrook. "Quantum generative adversarial learning". In: *Physical review letters* 121.4 (2018), p. 040502.

[9] Pierre-Luc Dallaire-Demers and Nathan Killoran. "Quantum generative adversarial networks". In: *Physical Review A* 98.1 (2018), p. 012324.

[10] Shu Lok Tsang et al. "Hybrid Quantum–Classical Generative Adversarial Network for High-Resolution Image Generation". In: *IEEE Transactions on Quantum Engineering* 4 (2022), pp. 1–19. URL: https://api.semanticscholar.org/CorpusID:254973945.

[11] Ling Hu et al. "Quantum generative adversarial learning in a superconducting quantum circuit". In: *Science advances* 5.1 (2019), eaav2761.

[12] He-Liang Huang et al. "Experimental quantum generative adversarial networks for image generation". In: *Physical Review Applied* 16.2 (2021), p. 024051.

[13] Daniel Silver et al. "Mosaiq: Quantum generative adversarial networks for image generation on nisq computers". In: *Proceedings of the IEEE/CVF International Conference on Computer Vision*. 2023, pp. 7030–7039.

[14] Su Yeon Chang et al. "Latent Style-based Quantum GAN for high-quality Image Generation". In: *arXiv e-prints*, arXiv:2406.02668 (June 2024), arXiv:2406.02668. DOI: 10.48550/arXiv.2406.02668. arXiv: 2406.02668 [quant-ph].

[15] Yixin Luo and Zhouwang Yang. "DynGAN: Solving Mode Collapse in GANs With Dynamic Clustering". In: *IEEE Transactions on Pattern Analysis and Machine Intelligence* 46.8 (2024), pp. 5493–5503. DOI: 10.1109/TPAMI.2024.3367532.

[16] Akash Srivastava et al. "Veegan: Reducing mode collapse in gans using implicit variational learning". In: *Advances in neural information processing systems* 30 (2017).

[17] Anders Boesen Lindbo Larsen et al. "Autoencoding beyond pixels using a learned similarity metric". In: *International conference on machine learning*. PMLR. 2016, pp. 1558–1566.

[18] Kaifeng Zhang. "On Mode Collapse in Generative Adversarial Networks". In: *Artificial Neural Networks and Machine Learning – ICANN 2021: 30th International Conference on Artificial Neural Networks, Bratislava, Slovakia, September 14–17, 2021, Proceedings, Part II*. Bratislava, Slovakia: Springer-Verlag, 2021, pp. 563–574. ISBN: 978-3-030-86339-5. DOI: 10.1007/978-3-030-86340-1_45. URL: https://doi.org/10.1007/978-3-030-86340-1_45.

[19] Shivani Tomar and Ankit Gupta. "A review on mode collapse reducing gans with gan's algorithm and theory". In: *GANs for Data Augmentation in Healthcare* (2023), pp. 21–40.

[20] Hoang Thanh-Tung and Truyen Tran. "Catastrophic forgetting and mode collapse in GANs". In: *2020 international joint conference on neural networks (ijcnn)*. IEEE. 2020, pp. 1–10.

[21] Martin Arjovsky, Soumith Chintala, and Léon Bottou. "Wasserstein Generative Adversarial Networks". In: *Proceedings of the 34th International Conference on Machine Learning*. Ed. by Doina Precup and Yee Whye Teh. Vol. 70. Proceedings of Machine Learning Research. PMLR, June 2017, pp. 214–223. URL: https://proceedings.mlr.press/v70/arjovsky17a.html.

[22] Martin Arjovsky, Soumith Chintala, and Léon Bottou. "Wasserstein Generative Adversarial Networks". In: *Proceedings of the 34th International Conference on Machine Learning*. Ed. by Doina Precup and Yee Whye Teh. Vol. 70. Proceedings of Machine Learning Research. PMLR, June 2017, pp. 214–223. URL: https://proceedings.mlr.press/v70/arjovsky17a.html.

[23] Ishaan Gulrajani et al. "Improved training of wasserstein GANs". In: *Proceedings of the 31st International Conference on Neural Information Processing Systems*. NIPS'17. Long Beach, California, USA: Curran Associates Inc., 2017, pp. 5769–5779. ISBN: 9781510860964.

[24] Diederik P. Kingma and Max Welling. "Auto-Encoding Variational Bayes". In: *2nd International Conference on Learning Representations, ICLR 2014, Banff, AB, Canada, April 14-16, 2014, Conference Track Proceedings*. 2014. arXiv: http://arxiv.org/abs/1312.6114v10 [stat.ML].

[25] Douglas A Reynolds et al. "Gaussian mixture models." In: *Encyclopedia of biometrics* 741.659-663 (2009), p. 3.

[26] Bing" "Xu et al. "Empirical evaluation of rectified activations in convolutional network". In: (2015).

[27] Kaiming He et al. "Delving Deep into Rectifiers: Surpassing Human-Level Performance on ImageNet Classification". In: *2015 IEEE International Conference on Computer Vision (ICCV)*. 2015, pp. 1026–1034. DOI: 10.1109/ICCV.2015.123.

[28] Adam Paszke et al. "Pytorch: An imperative style, high-performance deep learning library". In: *Advances in neural information processing systems* 32 (2019).

[29] William Falcon and The PyTorch Lightning team. *PyTorch Lightning*. Version 1.4. Mar. 2019. DOI: 10.5281/zenodo.3828935. URL: https://github.com/Lightning-AI/lightning.

[30] Ville Bergholm et al. "PennyLane: Automatic differentiation of hybrid quantum-classical computations". In: *arXiv e-prints*, arXiv:1811.04968 (Nov. 2018), arXiv:1811.04968. DOI: 10.48550/arXiv.1811.04968. arXiv: 1811.04968 [quant-ph].

[31] Christian Szegedy et al. "Rethinking the Inception Architecture for Computer Vision". In: *2016 IEEE Conference on Computer Vision and Pattern Recognition (CVPR)*. 2016, pp. 2818–2826. DOI: 10.1109/CVPR.2016.308.

[32] Francesca De Falco et al. "Towards efficient quantum hybrid diffusion models". In: *arXiv preprint arXiv:2402.16147* (2024).

[33] Martin Heusel et al. "Gans trained by a two time-scale update rule converge to a local nash equilibrium". In: *Advances in neural information processing systems* 30 (2017).

[34] Yahui Liu and Yang Li. *Metrics of GANs*. https://github.com/yhlleo/GAN-Metrics. Accessed: 2024-09-02, Online. 2021.

[35] Eitan Richardson and Yair Weiss. "On GANs and GMMs". In: *Proceedings of the 32nd International Conference on Neural Information Processing Systems*. NIPS'18. Montréal, Canada: Curran Associates Inc., 2018, pp. 5852–5863.

[36] Zhou Wang, Eero P. Simoncelli, and Alan C. Bovik. "Multi-scale structural similarity for image quality assessment". English (US). In: *Conference Record of the Asilomar Conference on Signals, Systems and Computers* 2 (2003). Conference Record of the Thirty-Seventh Asilomar Conference on Signals, Systems and Computers ; Conference date: 09-11-2003 Through 12-11-2003, pp. 1398–1402. ISSN: 1058-6393.

[37] Y. Lecun et al. "Gradient-based learning applied to document recognition". In: *Proceedings of the IEEE* 86.11 (1998), pp. 2278–2324. DOI: 10.1109/5.726791.

[38] Han Xiao, Kashif Rasul, and Roland Vollgraf. "Fashion-MNIST: a Novel Image Dataset for Benchmarking Machine Learning Algorithms". In: *arXiv e-prints*, arXiv:1708.07747 (Aug. 2017), arXiv:1708.07747. DOI: 10.48550/arXiv.1708.07747. arXiv: 1708.07747 [cs.LG].

[39] Diederik Kingma and Jimmy Ba. "Adam: A Method for Stochastic Optimization". In: *International Conference on Learning Representations (ICLR)*. San Diega, CA, USA, 2015.

[40] Andrew A Neath and Joseph E Cavanaugh. "The Bayesian information criterion: background, derivation, and applications". In: *Wiley Interdisciplinary Reviews: Computational Statistics* 4.2 (2012), pp. 199–203.

[41]  Amira Abbas et al. "The power of quantum neural networks". In: *Nature Computational Science* 1.6 (2021), pp. 403–409.

[42]  Laurens Van der Maaten and Geoffrey Hinton. "Visualizing data using t-SNE." In: *Journal of machine learning research* 9.11 (2008).

# Appendix

**Triple MNIST:** In Fig. 12 we plot the evaluation metrics calculated at the end of each epoch. The TL initialisation achieves the best FD, as well as higher SSIM and PSNR scores. This reflects improved perceptual fidelity, however, the cosine similarity metric remains largely unchanged across all variants. The latent space structure and generative diversity benefit from transfer learning, which also translates to better image quality.

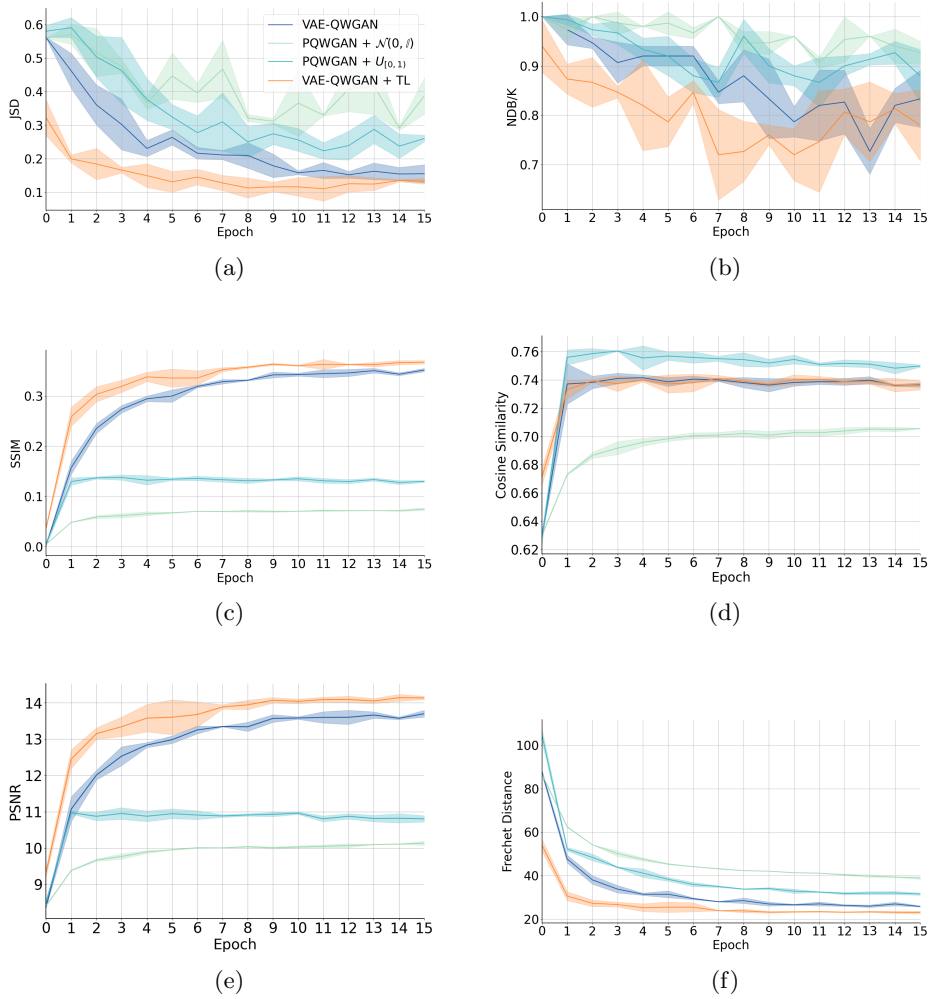

**Fig. 12**: Plots for Triple MNIST evaluating various metrics – **(a)** JSD **(b)** NDB **(c)** SSIM **(d)** Cosine Similarity **(e)** PSNR **(f)** FD – as a function of the training epochs for different benchmark quantum models.

**Triple Fashion MNIST:** In Fig. 13 we plot the evaluation metrics calculated at the end of each epoch. The TL initialisation again demonstrates improved convergence and better scores in several metrics. However, the difference between that and the non-TL initialisation is less pronounced than before. This shows how the increased difficulty of the dataset requires a smarter initialisation strategy.

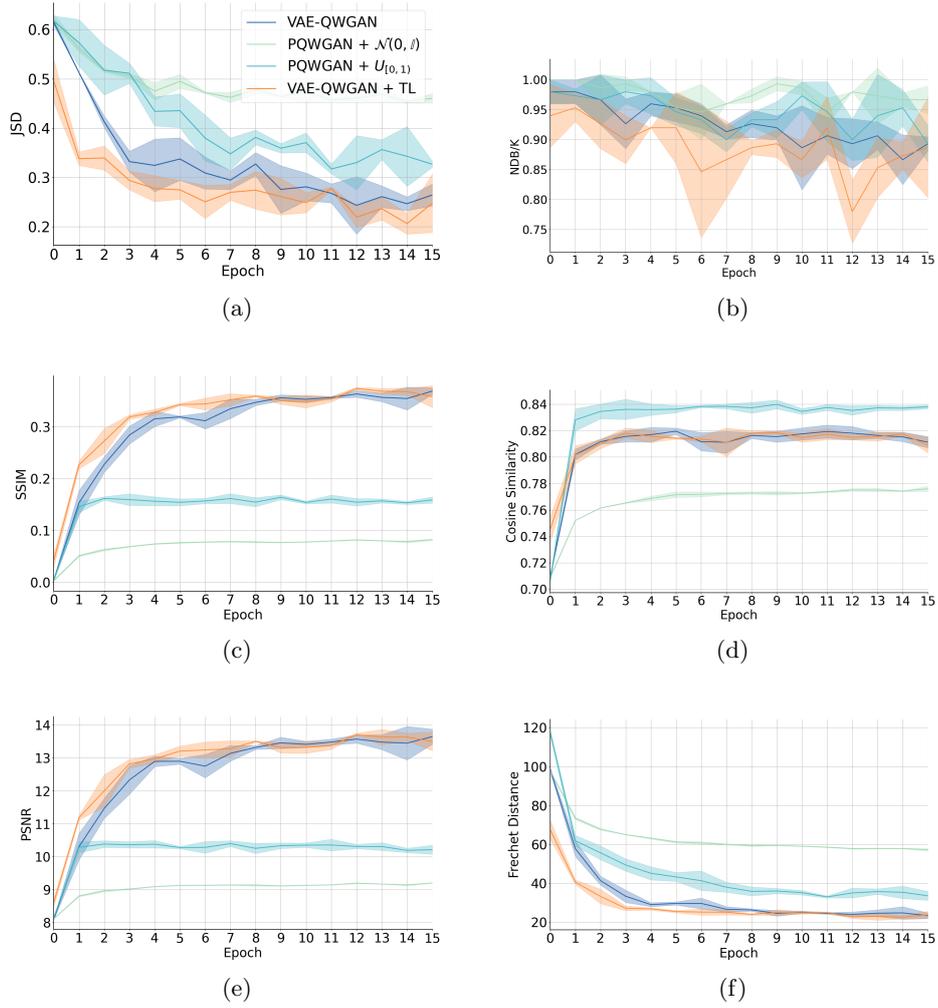

**Fig. 13**: Plots for Triple Fashion MNIST evaluating various metrics – **(a)** JSD **(b)** NDB **(c)** SSIM **(d)** Cosine Similarity **(e)** PSNR **(f)** FD – as a function of the training epochs for different benchmark quantum models.